\documentclass{iaus}
\usepackage{graphicx}

\title[Mixing in magnetic OB stars] %% give here short title %%
{Mixing in two magnetic OB stars discovered by the MiMeS collaboration}

\author[Thierry Morel]   %% give here short author list %%
{Thierry Morel}  

\affiliation{Institut d'Astrophysique et de G\'eophysique, Universit\'e de Li\`ege, 4000 Li\`ege, Belgium}

\pubyear{2008}
\volume{xxx}  %% insert here IAU Symposium No.
\pagerange{119--126}
\jname{Title of your IAU Symposium}
\editors{A.C. Editor, B.D. Editor \& C.E. Editor, eds.}
\begin{document}

\maketitle

\begin{abstract}
Recent observational and theoretical arguments suggest that magnetic OB stars may suffer more mixing than their non magnetic analogs. We present the results of an NLTE abundance study revealing a lack of CN-cycled material at the surface of two magnetic stars discovered by the MiMeS project (NGC2244 \#201 and HD 57682). The existence of a strong magnetic field is therefore not a sufficient condition for deep mixing in main-sequence OB stars. 
\keywords{stars: magnetic fields, stars: early-type, stars: abundances, stars: fundamental parameters, stars: individual (NGC2244 \#201, HD 57682)}
\end{abstract}

\firstsection % if your document starts with a section,

\section{Introduction}
The existence of a sizeable population of slowly-rotating and unevolved, yet N-rich, B stars in the Magellanic clouds (\cite[Hunter et al. 2008]{hunter}) and in the Galaxy (e.g., \cite[Gies \& Lambert 1992]{gies}) urges the need to (re)address the impact that magnetic fields may have on mixing of the internal layers in OB stars. This is mainly motivated by two considerations. First, the Geneva evolutionary models incorporating magnetic fields generated through dynamo action predict a greater amount of mixing and hence higher CNO abundance anomalies (\cite[Maeder \& Meynet 2005]{maeder}). Second, indication for a high incidence of a nitrogen excess in magnetic B stars was found by \cite[Morel et al. (2008)]{morel} who found 8 out of the 10 magnetic stars in their sample to be N rich by a factor $\sim$3. To further characterize the mixing properties of magnetic OB stars, we present the first results of an NLTE abundance study of a number of stars recently detected by the MiMeS collaboration. 

\section{Observations and results}
High-resolution ($R$ $\sim$ 46,000) FIES spectra of the four O9--B2 IV--V targets (NGC 2244 \#201, Par 1772, NU Ori and HD 57682) were obtained in late 2009 at the Nordic Optical Telescope. Spectropolarimetric observations of these stars revealed a field with a strength in the range 500--2000 G (\cite[Alecian et al. 2008]{alecian}; \cite[Grunhut et al. 2009]{grunhut}; \cite[Petit et al. 2008]{petit}). Spectral synthesis of HD 57682 using the NLTE, unified code CMFGEN provides $T_{\rm eff}$=34500$\pm$1000 K and $\log g$=4.0$\pm$0.2 dex (\cite[Grunhut et al. 2009]{grunhut}).

The atmospheric parameters are derived purely on spectroscopic grounds: $\log g$ from fitting the collisionally-broadened wings of the Balmer lines, $T_{\rm eff}$ from ionisation balance of various species (He, N, Ne and/or Si) and the microturbulence, $\xi$, from requiring the abundances yielded by the O II lines to be independent of their strength. The abundances are computed using Kurucz atmospheric models, the NLTE line-formation codes DETAIL/SURFACE and classical curve-of-growth techniques. Here we present the results for the two narrow-lined stars NGC 2244 \#201 and HD 57682. The two fast rotators remain to be analysed using spectral synthesis techniques. The atmospheric parameters and abundances are provided in Table 1 where they can be compared with the values obtained following exactly the same methodology for the magnetic, N-rich star $\tau$ Sco.
 
\begin{table}
\begin{center}
\caption{Atmospheric parameters and elemental abundances of NGC 2244 \#201 and HD 57682 (on a scale in which $\log \epsilon$[H]=12). Results obtained for $\tau$ Sco using exactly the same tools and techniques are shown for comparison (\cite[Hubrig et al. 2008]{hubrig}). The number of used lines is given in brackets. A blank indicates that no value could be determined. The solar [N/C] and [N/O] ratios are --0.60$\pm$0.08 and --0.86$\pm$0.08 dex, respectively (\cite[Asplund et al. 2009]{asplund}).}
\label{tab1}
%{\scriptsize
\begin{tabular}{l|cc|c}\hline 
                        & NGC 2244 \#201      & HD 57682             & $\tau$ Sco\\\hline
$T_{\rm eff}$ (K)       & 27000$\pm$1000      & 33000$\pm$1000       & 31500$\pm$1000\\
$\log g$                & 4.20$\pm$0.15       & 4.00$\pm$0.15        & 4.05$\pm$0.15\\
$\xi$ (km s$^{-1}$)     & 3$\pm$3             & 5$\pm$5$^{a}$        & 2$\pm$2\\ 
$v\sin i$ (km s$^{-1}$) & 22$\pm$2            & 25$\pm$4             & 8$\pm$2\\
He/H                    & 0.072$\pm$0.023 (9) & 0.106$\pm$0.030 (10) & 0.085$\pm$0.027 (9)\\
$\log \epsilon$(C)      & 8.22$\pm$0.13 (6)   & 8.20$\pm$0.19 (6)    & 8.19$\pm$0.14 (15)\\
$\log \epsilon$(N)      & 7.68$\pm$0.13 (20)  & 7.52$\pm$0.25 (8)    & 8.15$\pm$0.20 (35)\\
$\log \epsilon$(O)      & 8.63$\pm$0.18 (31)  & 8.31$\pm$0.21 (14)   & 8.62$\pm$0.20 (42)\\
$\log \epsilon$(Ne)     & 8.02$\pm$0.12 (7)   & 7.95$\pm$0.17 (1)    & 7.97$\pm$0.10 (5)$^{b}$\\
$\log \epsilon$(Mg)     & 7.29$\pm$0.20 (1)   & 7.37$\pm$0.18 (1)    & 7.45$\pm$0.09 (2)\\
$\log \epsilon$(Al)     & 6.20$\pm$0.13 (3)   & 6.07$\pm$0.21 (1)    & 6.31$\pm$0.29 (3)\\
$\log \epsilon$(Si)     & 7.41$\pm$0.25 (5)   & 7.47$\pm$0.32 (5)    & 7.24$\pm$0.14 (9)\\
$\log \epsilon$(S)      & 7.30$\pm$0.19 (1)   &                      & 7.18$\pm$0.28 (3)\\ 
$\log \epsilon$(Fe)     & 7.33$\pm$0.13 (20)  &                      & 7.33$\pm$0.31 (13)\\
${\rm [N/C]}$           & --0.54$\pm$0.14     & --0.68$\pm$0.30      & --0.04$\pm$0.25\\
${\rm [N/O]}$           & --0.95$\pm$0.21     & --0.79$\pm$0.19      & --0.47$\pm$0.29\\\hline
\end{tabular}
%}
\end{center}
\vspace{1mm}
\scriptsize{
{\it Notes:}\\
$^a$ Assumed values.\\
$^b$ From Morel \& Butler (2008).}
\end{table}

\section{Discussion}
These two main sequence stars do not show evidence for contamination of their surface layers by core-processed material. In the case of NGC 2244 \#201, the CNO logarithmic abundance ratios are consistent with the solar values, fully confirming the results of \cite[Vrancken et al. (1997)]{vrancken}. The results for HD 57682 are more uncertain owing to the weakness of the spectral lines and their strong $T_{\rm eff}$ sensitivity, but there is no indication for significant departures from the solar ratios either. It thus appears that these two stars do not display the N excess observed in other magnetic B stars (\cite[Morel et al. 2008]{morel}). We conclude that the relationship between magnetic fields and an N excess may only be statistical and that other (still elusive) parameters may control the amount of mixing experienced by B-type stars. This is particularly well illustrated by considering the dichotomy between the CNO abundance ratios of HD 57682 and $\tau$ Sco (Table \ref{tab1}) despite having similar characteristics in terms of location in the HR diagram and rotation rate.

%\begin{discussion}
%\end{discussion}

\end{document}